\documentclass[twocolumn]{autart}
\usepackage{mathtools}
\usepackage{amssymb}
\usepackage{balance}
\usepackage{enumitem}
\usepackage{xfrac}
\usepackage{array}
\usepackage{bm}
\usepackage{algorithm}
\usepackage{algorithmicx}
\usepackage{algpseudocode}

\usepackage{color}
\begin{document}
\begin{frontmatter}
\title{Consistent identification of continuous-time systems under multisine input signal excitation\thanksref{footnoteinfo}} 

\thanks[footnoteinfo]{This paper was not presented at any IFAC 
meeting. Corresponding author: R.~A.~Gonz\'alez.}

\author[KTH]{Rodrigo A. Gonz\'alez}\ead{grodrigo@kth.se},    % Add the 
\author[KTH]{Cristian R. Rojas}\ead{crro@kth.se},    % Add the 
\author[UON]{Siqi Pan}\ead{siqi.pan@uon.edu.au},               % e-mail address 
\author[UON]{James S. Welsh}\ead{james.welsh@newcastle.edu.au}  % (ead) as shown

\address[KTH]{Division of Decision and Control Systems, KTH Royal Institute of Technology, 10044 Stockholm, Sweden}  % Please supply                                              
\address[UON]{School of Electrical Engineering and Computing, University of Newcastle, Callaghan, 2308 NSW, Australia} 


\begin{abstract}
For many years, the Simplified Refined Instrumental Variable method for Continuous-time systems (SRIVC) has been widely used for identification. The intersample behaviour of the input plays an important role in this method, and it has been shown recently that the SRIVC estimator is not consistent if an incorrect assumption on the intersample behaviour is considered. In this paper, we present an extension of the SRIVC algorithm that is able to deal with continuous-time multisine signals, which cannot be interpolated exactly through hold reconstructions. The proposed estimator is generically consistent for any input reconstructed through zero or first-order-hold devices, and we show that it is generically consistent for continuous-time multisine inputs as well. The statistical performance of the proposed estimator is compared to the standard SRIVC estimator through extensive simulations.
\end{abstract}
\end{frontmatter}

\section{Introduction}
System identification involves using measured input and output data for building mathematical models that characterise a system's behaviour. Different approaches to system identification have been developed depending on whether a discrete-time (DT) or continuous-time (CT) model is needed. Continuous-time system identification has applications in many areas of science and engineering such as economics, biology, physics and control, with comprehensive literature written on the subject \cite{rao2006identification,Garnier2008book,young2012recursive}. Although the system identification community has focused mainly in DT setups, as it has been investigated during a predominantly digital era, there are many reasons why CT system identification has had a renewed interest during the last decades~\cite{garnier2015direct}. For example, model coefficients are directly linked to physical parameters, and more parsimonious models can be obtained as knowledge of the relative degree of the CT system can be accommodated. Also, contrary to DT system identification using the forward shift operator, irregular and fast sampling can be easily handled, since the associated parameters remain invariant with respect to the varying sampling period and the model poles do not become statistically ill-defined as the sampling period decays to zero.

One of the main difficulties in CT system identification is the treatment of time derivatives. Since the goal is to obtain an estimate of a CT system, knowledge of the derivatives of the input and output are, either explicitly or implicitly, required. However, these derivatives are not exactly computable when only sampled input-output data is obtained. To overcome this problem, many algorithms have been suggested (see, e.g., \cite{sinha1991identification,rao2006identification} and the references therein). One of the most popular algorithms is the Simplified Refined Instrumental Variable method for Continuous-time systems (SRIVC), which was first presented in \cite{young1980refined}. This method has been suggested for general use due to its robustness and accuracy in practical applications~\cite{garnier2014advantages}. Many further extensions of this method also exist in the literature, for example, to handle non-uniformly sampled data \cite{huselstein2002approach} or multi-input systems \cite{garnier2007optimal}. Extensions to output error (OE) and Box-Jenkins (BJ) models \cite{chen2013refined}, unification of DT and CT transfer function estimation~\cite{young2015refined}, and comprehensive consistency \cite{pan2020consistency} and asymptotic efficiency \cite{pan2020efficiency} analyses have also been presented.

The SRIVC algorithm uses interpolation of the input and output data in order to compute filtered regressor and instrument vectors in an iterative estimation procedure. This reconstruction of the CT input and output signals is usually implemented through simple interpolation schemes like zero-order hold (ZOH) or first-order hold (FOH) devices, independently of the nature of the true signals \cite{garnier2004time}. For inputs that can be described exactly with these reconstruction schemes, the SRIVC estimator has recently been shown to be generically consistent and asymptotically efficient \cite{pan2020consistency,pan2020efficiency}. However, when the intersample behaviour assumption on the model input does not match that of the system input, continuous-time estimation methods can deliver large estimation errors if the sampling period is large~\cite{schoukens1994identification}, and in particular, the SRIVC estimator is known to be generically inconsistent in this case. Important input signals for identification that cannot be described by holds are band-limited signals such as multisines. These signals are advantageous due to their flexibility regarding power spectrum design, time domain averaging possibilities, simplification of the model validation step and finite sample estimation performance \cite{schoukens1994advantages}. For these input signals, the complete CT input signal is known to the practitioner, but the SRIVC procedure only performs simple interpolations of the input, which impact its consistency regardless of the sampling period.

In summary, in this paper,
\vspace{-0.1cm}
	\begin{itemize}
		\item
		we present a refinement of the SRIVC method that is shown to yield generic consistency of the estimated model parameters for CT multisine input signal excitations;
		\item
		we prove that, given knowledge of the CT multisine input signal and measured output samples, the exact computation of the input regressors is necessary and sufficient for a generically consistent estimate of the CT system;
		\item
		we propose a computationally efficient algorithm for computing the regressors under the multisine case; and
		\item
		we exemplify the consistency of the proposed estimator through extensive Monte Carlo simulations.
	\end{itemize}
\vspace{-0.1cm}
The remainder of this paper is organised as follows. The identification problem is formulated in Section \ref{sec2}. Section \ref{sec3} provides a description of the SRIVC estimator and its consistency properties. The proposed SRIVC-type method is presented and analysed in Section \ref{sec4}, and Section \ref{sec5} illustrates this method with extensive numerical examples. Finally, conclusions are drawn in Section~\ref{sec6}.

%%%%%%%%%%%%%%%%%%%%%%%%%%%%%%%%%%%%%%%%%%%%%%%%%%%%%%%%%%%%%%%%%%%%%%%%%%%%%%%%%%%%%%%%%%%%%%%%%%%%%%%%%%%%%%%%%%%%%%%
\vspace{-0.25cm}
\section{Problem formulation}
\label{sec2}
Consider a linear and time-invariant (LTI), causal, stable, proper, single-input single-output, CT system
\begin{equation}
\label{eq1}
x(t) = \frac{B^*(p)}{A^*(p)} u(t), \notag 
\end{equation}
where $p$ is the Heaviside operator, i.e., $p g(t) := \text{d}g(t)/\text{d}t$, and the numerator and denominator polynomials are coprime and given by
\begin{align}
B^*(p) &= b_{m^*}^* p^{m^*} + b_{m^*-1}^* p^{{m^*}-1} + \dots + b_0^*, \notag \\
A^*(p) &= a_{n^*}^* p^{n^*} + a_{n^*-1}^* p^{{n^*}-1} + \dots + a_1^* p+1. \notag
\end{align}
Suppose that the CT input $u(t)$ is known from $t=t_1$ to $t=t_N$, where the sampling is regular in time unless explicity stated otherwise, and that $N$ noisy measurements of the output $x(t)$ are obtained at the instants $\{t_k\}_{k=1}^N$. In other words, the output observations are given by
\begin{equation}
\label{outputdescription}
y(t_k) =x(t_k) + v(t_k), \quad k=1,\dots,N, 
\end{equation}
where it is assumed that the sampled noise sequence $\{v(t_k)\}$ can be described as a zero-mean and finite variance random process. Due to the nature of the sampled signals and the difficulty of computing the time-derivative of CT white noise, which does not have finite variance \cite{aastrom2012introduction}, we only consider DT noise in this paper.

To identify the system, we propose the model structure
\begin{equation}
G(p)= \frac{b_m p^{m} + b_{m-1} p^{m-1} + \dots + b_0}{a_n p^{n} + a_{n-1} p^{n-1} + \dots + a_1 p+1},  \notag
\end{equation}
where the parameter vector 
\begin{equation}
\bm{\theta} := \begin{bmatrix}
a_1, & a_2, &\dots, & a_n, & b_0, & b_1, &\dots, & b_m 
\end{bmatrix}^\top \notag
\end{equation}
needs to be estimated. The goal is to obtain an accurate model of the CT system $G^*(p):=B^*(p)/A^*(p)$ given the knowledge of $N$ samples of the output measurements and the CT input signal. Note that in this framework the input signal is not limited to hold reconstructions. Hence, the description includes the standard framework where $u(t)$ is assumed to be obtained through a ZOH or FOH and extends to more general inputs, such as continuous-time multisines \cite{schoukens1994advantages}.

The identification of the system $G^*(p)$ can be done by obtaining the data points $\{u(t_k),y(t_k)\}$ and applying a method for CT system identification, such as in \cite{chen2013refined}, or as in  \cite{young1980refined,maruta2013projection,gonzalez2018asymptotically} for regular sampling schemes. In most of these algorithms, however, the hold reconstructions of the input and output are assumed, and they are independent of the exact nature of the signals. In this work, we show that the knowledge of the exact intersample behaviour of the input can provide further insights for a better design of the identification procedure. 

%%%%%%%%%%%%%%%%%%%%%%%%%%%%%%%%%%%%%%%%%%%%%%%%%%%%%%%%%%%%%%%%%%%%%%%%%%%%%%%%%%%%%%%%%%%%%%%%%%%%%%%%%%%%%%%%%%%%%%%

\section{The Simplified Refined Instrumental Variable method for Continuous-time systems (SRIVC)}
\label{sec3}
The SRIVC estimator is an adaptive instrumental variable algorithm where parameter-dependent CT filters are updated iteratively. In each step, the instruments are computed using the parameter estimate obtained in the previous iteration until the model parameters have converged. The iterative procedure of the SRIVC algorithm is designed so that the sum of squares of the residuals (also called the generalised equation errors or GEEs) $\varepsilon(t_k)$, is minimised. The residuals are written as
	\begin{align}
	\varepsilon(t_k) :&= y(t_k)-\frac{B(p)}{A(p)}u(t_k) \notag \\
	\label{gee}
	&= A(p)y_f(t_k)-B(p)u_f(t_k),  
	\end{align}
where
\begin{equation}
y_f(t_k) = \frac{1}{A(p)}y(t_k) \textnormal{, and } u_f(t_k) = \frac{1}{A(p)}u(t_k). \label{filt1}
\end{equation}
Note that in \eqref{gee} and \eqref{filt1} we have adopted a mixed notation of CT operators and DT data. Since this dichotomy is repeatedly encountered in this paper, we formalise it in the following remark.
\begin{rem}\hspace{-0.2cm}\textbf{.}
	\label{remark1}
In this paper, $G(p)x(t_k)$ means that the DT signal $x(t_k)$ is interpolated in some manner, e.g., using a ZOH or FOH, and the resultant output through the CT filter $G(p)$ is sampled at $t=t_k$. On the other hand, $\{G(p)x(t)\}_{t_k}$ (or $[G(p)x(t)]_{t_k}$ in the vector-valued case) means that the CT signal $x(t)$ is filtered through $G(p)$, and later sampled at $t=t_k$. 
\end{rem}
The SRIVC method is described in Algorithm \ref{algorithm1}, where we denote $\bm{\varphi}_f(t_k)$ as the filtered regressor vector, $\hat{\bm{\varphi}}_f(t_k)$ as the filtered instrument vector, and $y_f(t_k)$ as the filtered output. Note that line 8 of Algorithm \ref{algorithm1} requires the DT signals to be prefiltered by CT transfer functions. This is usually done by assuming a ZOH or FOH reconstruction for the input and output signals and then simulating the response by using, for example, the \texttt{lsim} command in MATLAB. Although this approach has provided a quick procedure to compute the filtered regressor and instrument vectors, it is prone to approximation errors that can jeopardise the statistical properties of the method.

\begin{algorithm}
%	\label{algorithm1}
	\renewcommand{\thealgorithm}{1}
	\caption{\hspace{-0.1cm}: SRIVC}
	\begin{algorithmic}[1]
		\State Input:  $\{u(t_k),y(t_k)\}_{k=1}^N$, model order $(n,m)$, initial vector estimate $\bm{\theta}_1$, tolerance $\epsilon$ and maximum number of iterations $M$
		\State Using $\bm{\theta}_1$, form the estimated system polynomials $A_1(p)$ and $B_1(p)$
		\State $j\gets 1$, $\textnormal{flag}\gets 1$
        \While{$\textnormal{flag}=1$ and $j\leq M$}
		\State Prefilter the (DT) input $\{u(t_k)\}_{k=1}^N$ and output  \hspace*{0.4cm} $\{y(t_k)\}_{k=1}^N$ to form
		\begin{flalign}
		\bm{\varphi}_f(t_k) &\gets \frac{1}{A_j(p)} \big[ -p y(t_k), \hspace{0.1cm} \dots,\hspace{0.1cm} - p^n y(t_k),& \notag \\
		\label{filteredregressor}
		&\hspace{2.4cm} u(t_k), \hspace{0.1cm}\dots,\hspace{0.1cm} p^m u(t_k)\big]^\top,& \\
		\hat{\bm{\varphi}}_f(t_k) &\gets \frac{1}{A_j(p)} \bigg[ -\frac{p B_j(p)}{A_j(p)} u(t_k), \hspace{0.1cm} \dots, & \notag \\
		\label{filteredinstrument}
		&\hspace{-0.2cm}-\frac{p^n B_j(p)}{A_j(p)} u(t_k), \hspace{0.1cm} u(t_k), \hspace{0.08cm} \dots, \hspace{0.08cm}p^m u(t_k)\bigg]^{\hspace{-0.05cm}\top}\hspace{-0.12cm},& \\
		\label{filteredoutput}
		y_f(t_k) &\gets \frac{1}{A_j(p)}y(t_k)& 
		\end{flalign} 
		\State 	Compute the parameter estimate		
		\begin{equation}
		\label{iterations}
		\bm{\theta}_{j+1} \gets \hspace{-0.05cm}\left[\sum_{k=1}^N  \hat{\bm{\varphi}}_f(t_k) \bm{\varphi}_f^\top(t_k)\right]^{-1}\hspace{-0.1cm}\left[ \sum_{k=1}^N \hat{\bm{\varphi}}_f(t_k) y_f(t_k)  \right]
		\end{equation}
		\If{$B_j(p)/A_j(p)$ is unstable}
		\State Reflect the unstable poles of $1/A_j(s)$ into the \hspace*{1cm} stable region of the complex $s$-plane
		\EndIf 
	    \If{$\dfrac{\|\bm{\theta}_{j+1}-\bm{\theta}_j\|}{\|\bm{\theta}_j\|}<\epsilon$}
     	\State $\textnormal{flag} \gets 0$
	    \EndIf 
		\State $j \gets j+1$
		\EndWhile{}
		\State Output: $\bm{\theta}_{j}$ and its associated model $B_j(p)/A_j(p)$.
	\end{algorithmic}
	\label{algorithm1}
\end{algorithm}

\begin{rem}\hspace{-0.2cm}\textbf{.}
In the SRIVC method, the user has several choices regarding the intersample behaviour assumptions. In particular, the intersample behaviour of the input in both \eqref{filteredregressor} and \eqref{filteredinstrument} can be chosen, as well as the reconstruction of the output signal for the filtering steps in~\eqref{filteredregressor} and \eqref{filteredoutput}. Usually the output is selected to have a FOH behaviour, since it is argued that it typically gives rise to a satisfactory approximation if the sampling period is small \cite{chen2017robust}.
\end{rem}

\subsection{Consistency Analysis of the SRIVC estimator}
Previous works \cite{young2008refined,young2015refined} have suggested that the SRIVC estimator uses the optimal instrumental variable terms, and that it minimises the prediction error and maximises the likelihood function, but they lack rigorous theoretical analysis regarding the influence of the interpolation of the input and output for the prefiltering step. Only recently \cite{pan2020consistency} has the intersample behaviour of the signals been taken into account for the consistency analysis. In \cite[Theorem 1]{pan2020consistency}, the generic consistency of the SRIVC estimator was proven for inputs that can be exactly interpolated by FOH or ZOH devices. More precisely, under mild assumptions regarding the sampling period and persistence of excitation of the input, the following statements are true for an input that is exactly reconstructible with FOH or ZOH interpolation:
	\begin{enumerate}[label={\arabic*}.]
		\item
		The matrix $\mathbb{E}\{\hat{\bm{\varphi}}_f(t_k)\bm{\varphi}_f^\top(t_k)\}$ is generically non-singular\footnote{In this context, generically non-singular means that the set $M=\{\bm{\theta}_j\in \mathbb{R}^n\colon A_j(p)$ is a stable polynomial, $\mathbb{E}\{\hat{\bm{\varphi}}_f(t_k)\bm{\varphi}_f^\top(t_k)\}$ is singular$\}$ has Lebesgue measure zero in $\mathbb{R}^n$.}\hspace{-0.05cm}.
		\item
		The true parameter $\bm{\theta}^*$ is the unique converging point.
		\item
		As the sample size $N$ approaches infinity, $\bm{\theta}_{j+1}$ in~\eqref{iterations} converges to $\bm{\theta}^*$ for $j\geq 1$.
	\end{enumerate}
Also, the effect of choosing a different intersample behaviour than that of the system input was also analysed in \cite[Corollary 3]{pan2020consistency}. In the following, we say that a \textit{correct specification of the intersample behaviour in the model input} occurs whenever the intersample behaviour of such signal in the SRIVC algorithm matches that of the input applied to the continuous-time system. Otherwise, we say that the intersample behaviour in the model input signal is misspecified. In~\cite{pan2020consistency} it was shown that the SRIVC estimator
\begin{enumerate}[label={\arabic*}.]
	\item remains generically consistent if a misspecification of the intersample behaviour is used for generating the filtered signals in the instrument vector $\hat{\bm{\varphi}}_f(t_k)$, and
	\item is generically inconsistent if a misspecification of the intersample behaviour is used for filtering the input signal in the regressor vector $\bm{\varphi}_f(t_k)$.
	\end{enumerate} 
This result indicates that the intersample behaviour of the input signal needs to be correctly taken into account for the consistency of the SRIVC estimator. In particular, it implies that if the system input is a signal that is not produced by a hold mechanism, the estimator will be generically inconsistent. This argument holds regardless of whether the additive noise $v(t_k)$ is white or coloured.

\section{Consistent SRIVC-type method}
\label{sec4}
As mentioned in the previous section, a correct specification of the intersample behaviour of the input (ZOH of FOH) in the regressor vector $\bm{\varphi}_f(t_k)$ guarantees generic consistency under mild conditions. The extension of this principle constitutes our main contribution. In this work, we propose an extension of the SRIVC method that computes the filtered regressors exactly for multisine input excitations, and prove its generic consistency.

The generalised equation error for the proposed approach is
	\begin{equation}
	\label{gee2}
	\varepsilon(t_k) = A(p)y_f(t_k)-\{B(p)u_f(t)\}_{t_k},
	\end{equation}
where $u_f(t) = \frac{1}{A(p)}u(t)$. %, and the notation $\{g(t)\}|_{t=t_k}$ means that the continuous-time signal $g(t)$ is evaluated at $t=t_k$. 
In \eqref{gee2}, the predicted output measurement is explicitly calculated by first computing the underlying CT signal, and later evaluating it at $t=t_k$. The proposed estimator follows the procedure described in Algorithm \ref{algorithm1}, but the filtered regressor and instrument vectors in Equations \eqref{filteredregressor} and \eqref{filteredinstrument} now become
\begin{flalign}
&\bm{\varphi}_f(t_k) = \Bigg[ \frac{-p}{A_j(p)} y(t_k), \hspace{0.1cm}\dots, \hspace{0.1cm}\frac{-p^n}{A_j(p)} y(t_k), \notag \\
\label{eval1}
& \hspace{1.1cm} \left\{\frac{1}{A_j(p)}u(t)\right\}_{t_k}\hspace{-0.05cm}, \hspace{0.1cm}\dots, \hspace{0.1cm}\left\{\frac{p^m}{A_j(p)} u(t)\right\}_{t_k}\Bigg]^\top,&
\end{flalign}
and
\begin{flalign}
&\hat{\bm{\varphi}}_f(t_k) = \Bigg[ -\frac{pB_j(p)}{A_j^2(p)} u(t), \hspace{0.1cm}\dots, \hspace{0.1cm}-\frac{p^n B_j(p)}{A_j^2(p)} u(t), & \notag \\
\label{eval2}
&\hspace{2.6cm} \frac{1}{A_j(p)}u(t),\hspace{0.1cm}\dots, \hspace{0.1cm}\frac{p^m}{A_j(p)}u(t)\Bigg]^\top_{t_k}.&
\end{flalign}
Note that the $t_k$ in \eqref{eval2} follows the notation in Remark~\ref{remark1}. 

\begin{rem}\hspace{-0.2cm}\textbf{.}
The proposed estimator is an extension of the standard SRIVC estimator, and uses the complete CT input signal for identification. For input signals that are reconstructed exactly through a ZOH or FOH (e.g., a PRBS signal), this estimator is equivalent to the SRIVC estimator. Thus, the SRIVC-type estimator with prefiltering stage given by~\eqref{eval1} and \eqref{eval2} is generically consistent under the same assumptions as in \cite{pan2020consistency} for ZOH and FOH inputs.  
\end{rem}
In order to further analyse the asymptotic properties of the proposed estimator, we now study its consistency for multisine inputs. %Later, an extension for arbitrary input excitations is presented.

\subsection{Consistency analysis for multisine inputs}
We consider multisine input signals of the form
\begin{equation}
\label{inputmultisine}
u(t)=\alpha_0+\sum_{l=1}^{m_u} \alpha_{l} \cos(\omega_l t + \psi_l),
\end{equation}
where $m_u,\{\alpha_l\}_{l=0}^{m_u},\{\omega_l\}_{l=1}^{m_u}$ and $\{\psi_l\}_{l=1}^{m_u}$ are input parameters. The frequencies $\omega_l$ are assumed to be positive and distinct, and without loss of generality we assume that the weights $\alpha_l$ are positive as well. It is well known that the output  in steady state of an asymptotically stable LTI filter $H(s)$ when $u(t)$ is applied is also a multisine, given by 
\begin{equation}
\label{eval3}
y(t)=H(0)\alpha_0+\sum_{l=1}^{m_u} \alpha_l|H(i\omega_l)| \cos(\omega_l t + \psi_l+\angle H(i\omega_l)).
\end{equation}
This property of LTI systems provides a natural way to obtain exact values for the signal evaluations in \eqref{eval1} and~\eqref{eval2}, and it is of low computational cost, since the prefiltering is directly obtained by evaluating \eqref{eval3} with the corresponding filter. Another advantage of this approach is that it extends naturally to non-uniformly sampled data. For such type of sampling, the proposed method is not as computationally intensive as the standard SRIVC method, since the algorithm only requires approximations of the filtered output $p^i A_j^{-1}(p)y(t_k), i=0,\dots,n$, instead of computing approximations of the filtered values of both $u(t_k)$ and $y(t_k)$. The filtered output computations can be carried out by, e.g., an adaptive Runge-Kutta method (as in~\cite{chen2013refined}), or by any oversampling technique with intersample behaviour assumptions.

We now prove the consistency of the proposed estimator for the multisine input. The assumptions we use during the analysis are the following:

\begin{enumerate}[label=(A{\arabic*})]
\item
\label{assumption1}
The true system $B^*(p)/A^*(p)$ is proper ($n^* \geq m^*$) and asymptotically stable with $A^*(p)$ and $B^*(p)$ being coprime.
\item
\label{assumption2}
The disturbance sequence $\{v(t_k)\}$ is a zero-mean stationary random process.
\item
\label{assumption3}
The number of sinusoids of the input, $m_u$, satisfies $m_u\geq (n+m)/2$, and the input offset, $\alpha_0$, is different from zero.
\item
\label{assumption4}
All the zeros of $A_j(p)$ have strictly negative real parts, $n\geq m$, with $A_j(p)$ and $B_j(p)$ being coprime.
\item
\label{assumption5}
The degrees of the polynomials in the model satisfy $\min(n-n^*,m-m^*)=0$.% \textcolor{blue}{and in case the denominator is overparameterised, $n-n^*\leq n^*-m^*$ holds.}
\end{enumerate}
Assumptions \ref{assumption1} and \ref{assumption2} are standard. The condition in Assumption \ref{assumption3} is a persistence of excitation requirement, where $\alpha_0 \neq 0$ is set only for simplicity in our derivations and can be removed\footnote{If no offset is considered, then at least $(n+m+1)/2$ sinusoids are required for our results.}\hspace{-0.05cm}. Given that the poles of unstable models are reflected in line 6 of Algorithm \ref{algorithm1}, Assumption \ref{assumption4} is met in practice. Assumption \ref{assumption5} takes into account the model structure, as it ensures a unique solution of the model parameters to be obtained. %\textcolor{blue}{In addition, the condition over the denominator over-parametrisation is needed for the well-posedness of the matrix being inverted during the computation of the iterates.}

Since deterministic inputs will be considered in conjunction with stochastic noise processes, our analysis uses the standard definition of expectation for quasi-stationary signals \cite[pp. 34]{ljung1998system}, which is
\begin{equation}
\overline{\mathbb{E}}\{g(t)\} := \lim_{N\to \infty} \frac{1}{N} \sum_{t=1}^N \mathbb{E}\{g(t)\}. \notag
\end{equation}
\begin{thm}\hspace{-0.2cm}\textbf{.}
	\label{theorem1}
	Consider the SRIVC-type estimator with a fixed sampling period $h$ and filtered regressor and instrument vectors given by \eqref{eval1} and \eqref{eval2} respectively, and suppose that Assumptions \ref{assumption1} to \ref{assumption5} hold. Then, the following statements are true:
\begin{enumerate}[label={\arabic*}.]
	\item
	There exists a maximum sampling period $h^*>0$ such that, if $h\leq h^*$, the matrix $\overline{\mathbb{E}}\{\hat{\bm{\varphi}}_f(t_k)\bm{\varphi}_f^\top(t_k)\}$ is generically non-singular.
	\item
	If $h\leq h^*$ and the SRIVC-type iterations converge, then the true parameter $\bm{\theta}^*$ is the unique converging point.
	\item As the sample size $N$ approaches infinity, $\bm{\theta}_{j+1}$ converges to $\bm{\theta}^*$ for $j\geq 1$.
\end{enumerate}
\end{thm}

\begin{pf*}{Proof.}
	
\textit{Proof of Statement 1}. By substituting
	\begin{equation}
	y(t_k) = \left\{\frac{B^*(p)}{A^*(p)}u(t)\right\}_{t_k} + v(t_k) \notag 
	\end{equation}
	into \eqref{eval1}, we find that $\bm{\varphi}_f(t_k)=\bm{\varphi}_{f1}(t_k)-\mathbf{v}_f(t_k)$, where 
	\small 
	\begin{align}
	\hspace{-0.21cm}\bm{\varphi}_{f1}(t_k)\hspace{-0.05cm}:= \hspace{-0.12cm}\Bigg[\dfrac{-p}{A_j(p)}&\left\{ \dfrac{B^*(p)}{A^*(p)}u(t) \right\}_{t_k} \hspace{-0.16cm},\hspace{0.1cm} \dots, \dfrac{-p^n}{A_j(p)}\left\{ \dfrac{B^*(p)}{A^*(p)}u(t) \right\}_{t_k} \notag \\
	&\hspace{0.1cm} \left\{\dfrac{u(t)}{A_j(p)} \right\}_{t_k}  \hspace{-0.1cm}, \hspace{0.1cm} \dots,\hspace{0.1cm} \left\{\dfrac{p^m u(t)}{A_j(p)} \right\}_{t_k}
	\Bigg]^\top,
	\end{align}
	and
	\begin{equation}
\label{Vtk}
\mathbf{v}_f(t_k):= \begin{bmatrix}
\dfrac{p}{A_j(p)} v(t_k), & \hspace{0.05cm}\dots, & \hspace{0.1cm} \dfrac{p^n}{A_j(p)} v(t_k), & 0, & \hspace{0.05cm}\dots, & 0  
\end{bmatrix}^\top.
	\end{equation}
	\normalsize
	On the other hand, we also have 
	\begin{equation}
	\hat{\bm{\varphi}}_f(t_k)=\mathbf{S}(-B_j,A_j) \left[ \dfrac{\mathbf{u}_{du}(t)}{A_j^2(p)} \right]_{t_k}, \notag
	\end{equation}
	where 
	\begin{equation}
	\label{Udu}
	\mathbf{u}_{du}(t) := \begin{bmatrix}
	\dfrac{\textnormal{d}^{n+m}}{\textnormal{d}t^{n+m}}u(t), & \dfrac{\textnormal{d}^{n+m-1}}{\textnormal{d}t^{n+m-1}}u(t), & \hspace{0.1cm}\dots, & \hspace{0.1cm} u(t)
	\end{bmatrix}^\top \hspace{-0.1cm},
	\end{equation}
	and $\mathbf{S}(-B_j,A_j)$ is the Sylvester matrix associated with the polynomials $-B_j(p)$ and $A_j(p)$, whose non-singularity follows from the same analysis done in \cite{pan2020consistency}, where Assumption \ref{assumption4} is used. With this, we compute 
	\small
	\begin{align}
	\overline{\mathbb{E}}\{\hat{\bm{\varphi}}_f(t_k)\bm{\varphi}_f^\top(t_k)\} &= \mathbf{S}(-B_j,A_j)\underbrace{\overline{\mathbb{E}}\left\{\left[ \dfrac{\mathbf{u}_{du}(t)}{A_j^2(p)} \right]_{t_k}\bm{\varphi}_{f1}^\top(t_k) \right\}}_{=:\bm{\Phi}} \notag \\
	\label{Psidef}
	&\hspace{-1.5cm}-\mathbf{S}(-B_j,A_j)\underbrace{\overline{\mathbb{E}}\left\{\left[ \dfrac{\mathbf{u}_{du}(t)}{A_j^2(p)} \right]_{t_k}\mathbf{v}_f^\top(t_k)\right\}}_{=:\bm{\Psi}}.
	\end{align}
	\normalsize
	Thus, for showing that $\overline{\mathbb{E}}\{\hat{\bm{\varphi}}_f(t_k)\bm{\varphi}_f^\top(t_k)\}$ is generically non-singular for a small enough sampling period $h$, it is sufficient to show that $\bm{\Psi}=\mathbf{0}$ and $\bm{\Phi}$ is generically non-singular for a small enough $h$. The difference between the analysis in \cite[Theorem 1]{pan2020consistency} and the proof in the current paper is that the signals of interest are hybrid in nature: some are evaluations of CT signals, whereas others are DT signals interpolated with a reconstruction device, such as a FOH.
	
	The proof of $\bm{\Psi}=\mathbf{0}$ can be found in Lemma \ref{lemma1} in the Appendix. Regarding the invertibility of $\bm{\Phi}$, we will conveniently write $\bm{\varphi}_{f1}(t_k)$ as $\bm{\varphi}_{f2}(t_k)+\bm{\Delta}(t_k)$, where $\bm{\Delta}(t_k) \in \mathbb{R}^{n+m+1}$ has entries
	\begin{equation}
	\label{deltai}
	\bm{\Delta}_i(t_k)= \begin{cases}
	\left\{\frac{p^i}{A_j(p)}x(t)\right\}_{t_k}-\frac{p^i}{A_j(p)} x(t_k) &\hspace{-0.2cm},\hspace{0.07cm} i=1,\dots,n \\
	0 &\hspace{-2.2cm},\hspace{0.07cm} i=n+1,\dots,n+m+1,
	\end{cases}
	\end{equation}
	and
	\begin{align}
	\bm{\varphi}_{f2}(t_k) &= \Bigg[
	\dfrac{-p B^*(p)}{A_j(p) A^*(p)}u(t), \hspace{0.1cm}\dots, \hspace{0.1cm}\dfrac{-p^n B^*(p)}{A_j(p)A^*(p)}u(t), \notag \\
	&\hspace{2cm} \quad \dfrac{1}{A_j(p)}u(t),  \hspace{0.1cm}\dots, \hspace{0.1cm} \dfrac{p^m}{A_j(p)}u(t)
	\Bigg]^\top_{t_k} \notag \\
	&= \mathbf{S}(-B^*,A^*) \left[ \frac{\mathbf{u}_{du}(t)}{A_j(p)A^*(p)} \right]_{t_k}, \notag 
	\end{align}
	with $\mathbf{S}(-B^*,A^*)$ being the Sylvester matrix associated with the polynomials $-B^*(p)$ and $A^*(p)$, which is non-singular since $A^*(p)$ and $B^*(p)$ are coprime. Hence, we can write the expected value of interest as
	\begin{align}
	\overline{\mathbb{E}}\{\hat{\bm{\varphi}}_f(t_k)\bm{\varphi}_f^\top(t_k)\} &= \mathbf{S}(-B_j,A_j)\bm{\Phi}_1 \mathbf{S}^\top(-B^*,A^*) \notag \\
	\label{righthandside}
	&\hspace{-1.9cm}+\mathbf{S}(-B_j,A_j)\overline{\mathbb{E}}\left\{\left[ \dfrac{\mathbf{u}_{du}(t)}{A_j^2(p)} \right]_{t_k}\bm{\Delta}^\top(t_k)\right\},
	\end{align} 
    where
	\begin{equation}
	\bm{\Phi}_1 := \overline{\mathbb{E}}\left\{ \left[\frac{\mathbf{u}_{du}(t)}{A_j^2(p)}\right]_{t_k} \left[\frac{\mathbf{u}_{du}(t)}{A_j(p)A^*(p)} \right]_{t_k}^\top \right\}. \notag 
	\end{equation}
	It is shown in Lemma \ref{lemma2} in the Appendix that $\bm{\Phi}_1$ is generically non-singular, which means that the first summand of the right hand side of \eqref{righthandside} is generically non-singular.
	
	Finally, as $h$ tends to zero, the infinity norm of the difference between the direct evaluation of a CT signal and its interpolated counterpart also tends to zero. Thus, $\bm{\Delta}(t_k)\to 0$ as $h\to 0$. This, together with the fact that (generic) non-singularity of a matrix is preserved under small-enough matrix perturbations \cite[Chap. 6]{Horn2012}, leads to the first statement of the theorem.
	
	\textit{Statement 2}. Suppose that $\bar{\bm{\theta}}$ is a limiting point of the iteration in \eqref{iterations}, where $\bm{\varphi}_f(t_k)$ and $\hat{\bm{\varphi}}_f(t_k)$ are defined as in~\eqref{eval1} and \eqref{eval2} respectively, and the corresponding polynomials are denoted by $\bar{A}(p)$ and $\bar{B}(p)$. These polynomials are coprime  by Assumption \ref{assumption4}. The ergodic lemmas in \cite{soderstrom1975ergodicity} and \cite[Lemma A4.3]{soderstrom1983instrumental} permit us to write the iteration equation \eqref{iterations}, at the converging point and as $N$ tends to infinity, as
	\begin{equation}
	\label{eqlimit}
	\overline{\mathbb{E}}\{\hat{\bm{\varphi}}_f(t_k,\bar{\bm{\theta}})\bm{\varphi}_f^\top(t_k,\bar{\bm{\theta}})\}^{-1} \overline{\mathbb{E}}\{\hat{\bm{\varphi}}_f(t_k,\bar{\bm{\theta}})\varepsilon(t_k,\bar{\bm{\theta}})\}=\mathbf{0},
	\end{equation}
	where $\varepsilon(t_k,\bar{\bm{\theta}})$ is the GEE \eqref{gee2} evaluated at the converging point. Since the matrix inverse in \eqref{eqlimit} is assumed to be non-singular, the second expectation in \eqref{eqlimit} must be zero, i.e.,
	\begin{equation}
	\label{eqcon2}
	\overline{\mathbb{E}}\{\hat{\bm{\varphi}}_f(t_k,\bar{\bm{\theta}})\varepsilon(t_k,\bar{\bm{\theta}})\}=\mathbf{0}.
	\end{equation}
	
	Let $\bar{A}(p)B^*(p)-\bar{B}(p)A^*(p)=h_0p^r+h_1 p^{r-1}+\dots + h_r$, where $r = \max(n+m^*,n^*+m)=n+m$. Then, the GEE in \eqref{gee2} can be rearranged as
	\begin{align}
    \varepsilon(t_k,\bar{\bm{\theta}})&=\left\{\frac{\bar{A}(p)B^*(p)-\bar{B}(p)A^*(p)}{\bar{A}(p)A^*(p)}u(t)\right\}_{t_k}+v(t_k) \notag \\
    &= \left\{\frac{\mathbf{u}_{du}^\top(t)}{\bar{A}(p)A^*(p)}\mathbf{h} \right\}_{t_k} + v(t_k), \notag
    \end{align}
	where $\mathbf{h} = \begin{bmatrix}
	h_0, & h_1, & \hspace{0.08cm}\dots, & h_{n+m}\end{bmatrix}^\top$.
	Now, note that the instrument vector $\hat{\bm{\varphi}}_f(t_k)$ can be written as
	\begin{equation}
	\hat{\bm{\varphi}}_f(t_k) = \mathbf{S}(-\bar{B},\bar{A})\left[\frac{\mathbf{u}_{du}(t)}{\bar{A}^2(p)}\right]_{t_k}, \notag
	\end{equation}
	where  $\mathbf{S}(-\bar{B},\bar{A})$ is a Sylvester matrix associated with the polynomials $\bar{B}(p)$ and $\bar{A}(p)$, which again is non-singular. So, we can express \eqref{eqcon2} as
	\begin{align}
	&\mathbf{0} = \mathbf{S}(-\bar{B},\bar{A}) \underbrace{\overline{\mathbb{E}}\left\{\left[\frac{\mathbf{u}_{du}(t)}{\bar{A}^2(p)}\right]_{t_k} \left[\frac{\mathbf{u}_{du}(t)}{\bar{A}(p)A^*(p)}\right]_{t_k}^\top\right\}}_{:=\bar{\bm{\Phi}}} \mathbf{h} \notag \\
	\label{eqcon3}
	&\hspace{0.2cm}+ \mathbf{S}(-\bar{B},\bar{A})\underbrace{\overline{\mathbb{E}}\left\{\left[\frac{\mathbf{u}_{du}(t)}{\bar{A}^2(p)}\right]_{t_k}v(t_k) \right\}}_{:=\bar{\bm{\Psi}}}.
	\end{align}
	Following a similar approach as in Lemma \ref{lemma1}, we conclude that $\bar{\bm{\Psi}}=\mathbf{0}$, and by Lemma \ref{lemma2}, $\bar{\bm{\Phi}}$ is generically non-singular. Thus, for \eqref{eqcon3} to hold we need $\mathbf{h}=\mathbf{0}$, which implies that
	\begin{align}
	\bar{A}(p)B^*(p)&-\bar{B}(p)A^*(p)=0 \notag \\
	\implies \frac{\bar{B}(p)}{\bar{A}(p)}&= \frac{B^*(p)}{A^*(p)}, \notag 
	\end{align} 
	i.e., $\bm{\theta}^*$ is the unique limiting point.

\textit{Statement 3}. The proof follows from the analysis made for proving Statement 3 of Theorem 1 in~\cite{pan2020consistency}.  \hspace*{\fill} \qed
\end{pf*}

Note that if the commonly used FOH (or ZOH) were chosen as the intersample behaviour of the signals when discretising the prefilters, the reconstruction of $u(t)$ would suffer from high frequency distortion, which usually leads to inaccuracies in the computation of $\bm{\varphi}_f(t_k)$ and $\hat{\bm{\varphi}}_f(t_k)$. As stated next, only an inaccurate computation of the regressor vector $\bm{\varphi}_f(t_k)$ causes generic inconsistency of the proposed method under CT multisine input excitation.

\begin{cor}\hspace{-0.2cm}\textbf{.}
Assume that the intersample behaviour in the model input is misspecified, but nevertheless satisfies $G(p)u(t_k) = \{G(p)u(t)\}_{t_k}$ as $h\to 0$. The SRIVC-type estimator with filtered regressor and instrument vectors given by \eqref{eval1} and \eqref{eval2} respectively
	\begin{enumerate}[label={\arabic*}.]
	\item remains generically consistent if a misspecification of the intersample behaviour is used for generating the filtered signals in the instrument vector $\hat{\bm{\varphi}}_f(t_k)$, and
	\item is generically inconsistent if a misspecification of the intersample behaviour is used for filtering the input signal in the regressor vector $\bm{\varphi}_f(t_k)$.
\end{enumerate}
\end{cor}
\begin{pf*}{Proof.}

\textit{Statement 1}: The result follows from the same logic as in the proof in \cite[Corollary 3, Statement 1]{pan2020consistency}.
	
\textit{Statement 2}: Statement 1 of Theorem \ref{theorem1} still holds by following the same steps as before, but this time the vector $\bm{\Delta}(t_k)$ in \eqref{deltai} will also have non-zero elements in its bottom $m+1$ entries. Namely, the $i$-th component of $\bm{\Delta}(t_k)$, with $i=n+1,\dots,n+m+1$, is now
\begin{equation}
\bm{\Delta}_i(t_k) = \frac{p^{i-n-1}}{A_j(p)}u(t_k)-\left\{\frac{p^{i-n-1}}{A_j(p)}u(t)\right\}_{t_k}, \notag 
\end{equation}
which still satisfies $\bm{\Delta}_i(t_k)\to 0$ as $h\to 0$. Thus, Theorem~\ref{theorem1} is valid for this case as well. However, Statement~2 of Theorem \ref{theorem1} does not yet hold. This fact follows from the same analysis done in the proof in \cite[Corollary 3, Statement 2]{pan2020consistency}.  \hspace*{\fill} \qed
\end{pf*}

\begin{rem}\hspace{-0.2cm}\textbf{.}
A similar procedure to \eqref{eval1} and \eqref{eval2} could be proposed for the computation of $y_f(t_k)$ by exploiting the fact that the noiseless output also corresponds to a multisine (thus, a more adequate reconstruction scheme could be designed). However, Remark 5 of \cite{pan2020consistency} suggests that, as the number of iterations tends to infinity, the GEE at the converging point does not depend on the intersample behavior of the output. Thus, if the iterations converge, a more precise filtering of the output is not needed.
\end{rem}

\subsection{The SRIVC-c algorithm}
To finalise this section, in Algorithm \ref{algorithm2} we provide a pseudo-code for computing the proposed SRIVC-type estimator, which is from now on labelled SRIVC-c.

\begin{algorithm}
	\renewcommand{\thealgorithm}{2}
	\caption{\hspace{-0.1cm}: SRIVC-c}
	\begin{algorithmic}[1]
		\State Input:  $\{u(t)\}_{t\in [t_1,t_N]}$, $\{y(t_k)\}_{k=1}^N$, model order $(n,m)$, initial vector estimate $\bm{\theta}_1$, tolerance $\epsilon$ and maximum number of iterations $M$
		\State Using $\bm{\theta}_1$, form the estimated system polynomials $A_1(p)$ and $B_1(p)$
		\State $j\gets 1$, $\textnormal{flag}\gets 1$
		\While{$\textnormal{flag}=1$ and $j\leq M$}
		\State Compute $\bm{\varphi}_f(t_k)$ and $\hat{\bm{\varphi}}_f(t_k)$ by \eqref{eval1} and \eqref{eval2},  \hspace*{0.45cm}
		where direct evaluations are performed as in \eqref{eval3} 
		\State Compute the filtered output $y_f(t_k)$ by \eqref{filteredoutput}
		\State Compute the parameter estimate as in \eqref{iterations}
		\If{$B_j(p)/A_j(p)$ is unstable}
		\State Reflect the unstable poles of $1/A_j(s)$ into the \hspace*{1cm} stable region of the complex $s$-plane
		\EndIf 
		\If{$\dfrac{\|\bm{\theta}_{j+1}-\bm{\theta}_j\|}{\|\bm{\theta}_j\|}<\epsilon$}
		\State $\textnormal{flag} \gets 0$
		\EndIf 
		\State $j \gets j+1$
		\EndWhile{}
		\State Output: $\bm{\theta}_{j}$ and its associated model $B_j(p)/A_j(p)$.
	\end{algorithmic}
	\label{algorithm2}
\end{algorithm}

%%%%%%%%%%%%%%%%%%%%%%%%%%%%%%%%%%%%%%%%%%%%%%%%%%%%%%%%%%%%%%%%%%%%%%%%%%%%%%%%
\section{Simulation examples}
\label{sec5}
Via numerical simulations under several experimental conditions, we evaluate the consistency of the standard SRIVC method and the proposed SRIVC-c method. For a multisine input, we examine the consistency of both methods for different regular sampling periods and also for irregular sampling. For the following tests, we consider the system
\begin{equation}
\label{systemexample}
G^*(p) = \frac{1.25}{0.25p^2+0.7p+1},
\end{equation}
where the parameters of interest are $a_1^*=0.7$, \mbox{$a_2^*=0.25$}, and $b_0^*=1.25$. Regarding the implementation of the standard SRIVC method, we have used the \texttt{srivc} command from the CONTSID toolbox version 7.3 for MATLAB \cite{garnier2018contsid}, under default initialisation and tolerance settings. It was set to estimate the best model among the correct model structure with a FOH as the intersample behaviour.

\subsection{Regular sampling}
We first test if the algorithms provide consistent estimates of the parameter vector $[a_1, \hspace{0.1cm} a_2, \hspace{0.10cm} b_0]^\top$. The system in \eqref{systemexample} is excited with the CT input
\begin{equation}
u(t) = \sin(0.714t)+ \sin(1.428t)+\sin(2.142t). \notag 
\end{equation}
The noiseless output is computed analytically by assuming that it corresponds to the output of the system at the stationary regime, i.e.,
\begin{equation}
x(t) = \sum_{k=1}^3|G^*(i\omega_k)|\sin(\omega_k t+\angle G^*(i\omega_k)), \notag 
\end{equation}
where $(\omega_1,\omega_2,\omega_3) = (0.714, 1.428,2.142)[\textnormal{rad}/\textnormal{s}]$. This output is sampled at $h = 0.3[\textnormal{s}]$ and is contaminated with additive noise, which is set as an i.i.d. Gaussian white noise sequence of variance 0.1. Sixty different sample sizes are considered, ranging logarithmically from $N=100$ to $N=25500$, and 300 Monte Carlo runs are performed for each value of $N$.

\begin{figure}
	\centering{
		\includegraphics[width=0.48\textwidth]{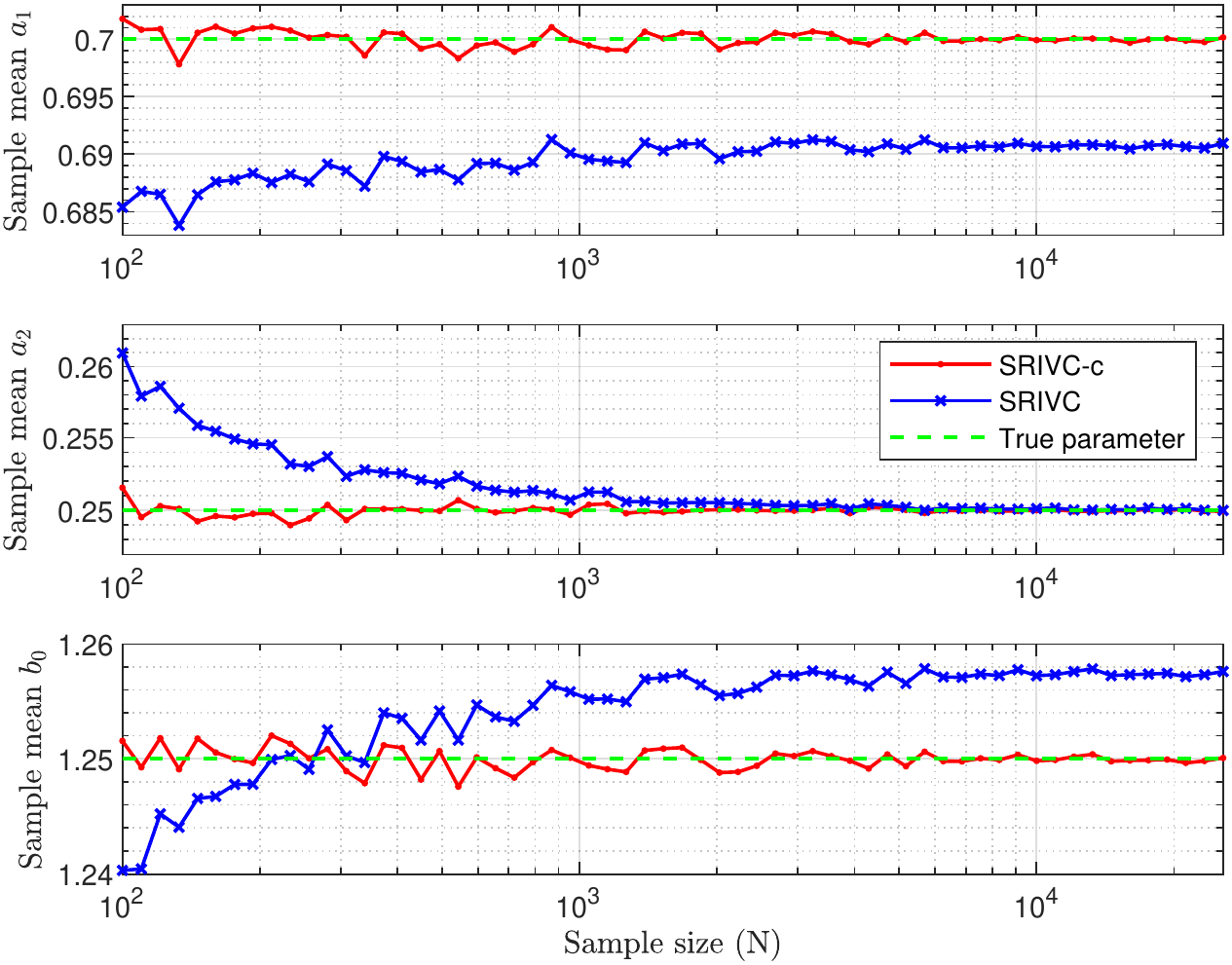}
		\caption{Regular sampling. Sample means of each estimated parameter for SRIVC-c (red), and standard SRIVC (blue). The true parameters are in dashed-green.}
		\label{fig1}}
\end{figure} 

\begin{figure}
	\centering{
		\includegraphics[width=0.48\textwidth]{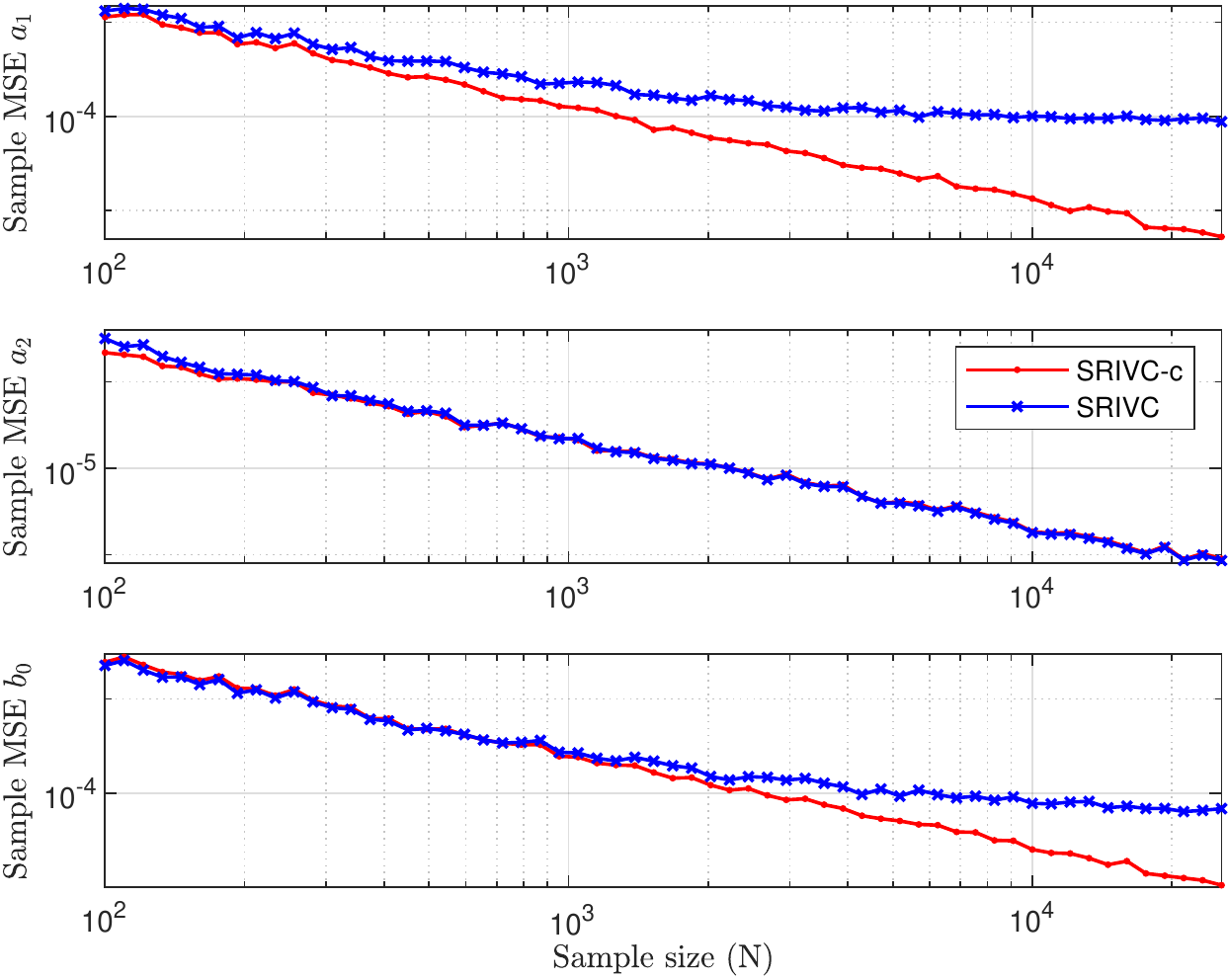}
		\caption{Regular sampling test. Sample MSE of each estimated parameter for SRIVC-c (red), and standard SRIVC (blue).}
		\label{fig2}}
\end{figure} 

Figures \ref{fig1} and \ref{fig2} show the sample means and sample mean square errors (MSEs) of each estimated parameter. The SRIVC-c estimator accurately identifies all parameters while the standard SRIVC method fails to recover the true parameter vector as $N$ increases. Regarding Figure \ref{fig2}, the MSEs for the SRIVC-c estimator decrease to zero. Together with the sampled means converging to the true values, these plots provide evidence for the consistency in mean square of the SRIVC-c estimator. Contrastingly, at least two out of the three estimated parameters given by the SRIVC method are biased, which empirically indicates that the SRIVC estimator is not consistent in this example.

\subsection{Different sampling periods}
We now study the effect of the intersample behaviour on the SRIVC-type estimates. Under the same input and noise variance as the previous simulation, we test the performance of each algorithm for a fixed number of output measurements ($N=2000$) with different regular sampling periods. Since the rise time of the system is approximately 2 seconds, a good choice for the sampling period should be between 0.2 and 0.5 seconds according to the criterion suggested in \cite{astroem1984computer}. In order to cover fast, normal and slow sampling, we test with sampling periods $h=0.06,0.2$ and $0.6$[s]. 

The sample mean and mean square error of each parameter over $300$ Monte Carlo runs for each sampling period are shown in Table \ref{table1}. On average, the SRIVC-c estimator delivers the true values of every parameter for all sampling periods in this study, whereas the SRIVC estimator only performs well (but anyway has noticeable bias) when the sampling period is small. For $h=0.6[\textnormal{s}]$, the large sampling period exaggerates the interpolation error of the input signal in the standard SRIVC estimator, which severely degrades its performance. This is confirmed by the order of magnitude of difference in MSE of the parameters given by the two estimators.

\begin{table}
	\caption{Sample mean and MSE of each parameter, for SRIVC and SRIVC-c, when $h=0.06, 0.2$ and $0.6[\textnormal{s}]$.}
	\centering
	\scriptsize
	\label{table1}
	\begin{tabular}{|@{\hspace{0.1cm}}m{0.92cm}|m{0.96cm}|@{\hspace{0.1cm}}m{0.63cm}@{\hspace{0.25cm}}|m{1.13cm}|m{1.13cm}|m{1.13cm}|}
		\hline
		Method & \begin{tabular}[c]{@{}c@{}}Param.\\ (Value)\end{tabular} & Stats.                                             & \centering{\begin{tabular}[c]{@{}c@{}}$h=0.06$\end{tabular}}                                                            & $h=0.2$                                                             & $h=0.6$                                                             \\ \hline
		& $a_1 (0.7)$                                                     & \begin{tabular}[c]{@{}c@{}}Mean\\ MSE\end{tabular} & \begin{tabular}[c]{@{}c@{}}$0.697$\\ $7.0\cdot 10^{-5}$\end{tabular} & \begin{tabular}[c]{@{}c@{}}$0.694$\\ $9.3\cdot 10^{-5}$\end{tabular} & \begin{tabular}[c]{@{}c@{}}$0.668$\\ $1.1\cdot 10^{-3}$\end{tabular} \\ \cline{2-6} 
		SRIVC  & $a_2 (0.25)$                                                     & \begin{tabular}[c]{@{}c@{}}Mean\\ MSE\end{tabular} & \begin{tabular}[c]{@{}c@{}}$0.253$\\ $1.8\cdot 10^{-5}$\end{tabular} & \begin{tabular}[c]{@{}c@{}}$0.251$\\ $1.1\cdot 10^{-5}$\end{tabular}  & \begin{tabular}[c]{@{}c@{}}$0.248$\\ $1.6\cdot 10^{-5}$\end{tabular} \\ \cline{2-6} 
		& $b_0 (1.25)$                                                      & \begin{tabular}[c]{@{}c@{}}Mean\\ MSE\end{tabular} & \begin{tabular}[c]{@{}c@{}}$1.244$\\ $1.6\cdot 10^{-4}$\end{tabular}    & \begin{tabular}[c]{@{}c@{}}$1.251$\\ $1.2\cdot 10^{-4}$\end{tabular} & \begin{tabular}[c]{@{}c@{}}$1.286$\\ $1.5\cdot 10^{-3}$\end{tabular} \\ \hline
		& $a_1 (0.7)$                                                     & \begin{tabular}[c]{@{}c@{}}Mean\\ MSE\end{tabular} & \begin{tabular}[c]{@{}c@{}}$0.700$\\ $6.0\cdot 10^{-5}$\end{tabular} & \begin{tabular}[c]{@{}c@{}}$0.699$\\ $6.3\cdot 10^{-5}$\end{tabular} & \begin{tabular}[c]{@{}c@{}}$0.700$\\ $6.7\cdot 10^{-5}$\end{tabular} \\ \cline{2-6} 
		\centering{SRIVC-c} & \centering{ $a_2 (0.25)$}                                                     & \begin{tabular}[c]{@{}c@{}}Mean\\ MSE\end{tabular} & \begin{tabular}[c]{@{}c@{}}$0.250$\\ $1.1\cdot 10^{-5}$\end{tabular} & \begin{tabular}[c]{@{}c@{}}$0.250$\\ $1.1\cdot 10^{-5}$\end{tabular} & \begin{tabular}[c]{@{}c@{}}$0.250$\\ $1.2\cdot 10^{-5}$\end{tabular} \\ \cline{2-6} 
		& $b_0 (1.25)$                                                      & \begin{tabular}[c]{@{}c@{}}Mean\\ MSE\end{tabular} & \begin{tabular}[c]{@{}c@{}}$1.249$\\ $1.3\cdot 10^{-4}$\end{tabular} & \begin{tabular}[c]{@{}c@{}}$1.250$\\ $1.3\cdot 10^{-4}$\end{tabular} & \begin{tabular}[c]{@{}c@{}}$1.250$\\ $1.3\cdot 10^{-4}$\end{tabular} \\ \hline
	\end{tabular}
\end{table}

\subsection{Irregular sampling}
We consider the same system described before, with the same input and noise variance. In this simulation study, 2000 irregularly sampled output measurements are obtained. The sampling interval is distributed uniformly between $h_{lb}$ and $h_{hb}$, where the lower bound is fixed at $h_{lb}=0.05$, while the upper bound is varied from $0.1$ to $0.6$. A total of 6 Monte Carlo simulations are performed with each simulation containing 300 runs.

\begin{figure}
	\centering{
		\includegraphics[width=0.475\textwidth]{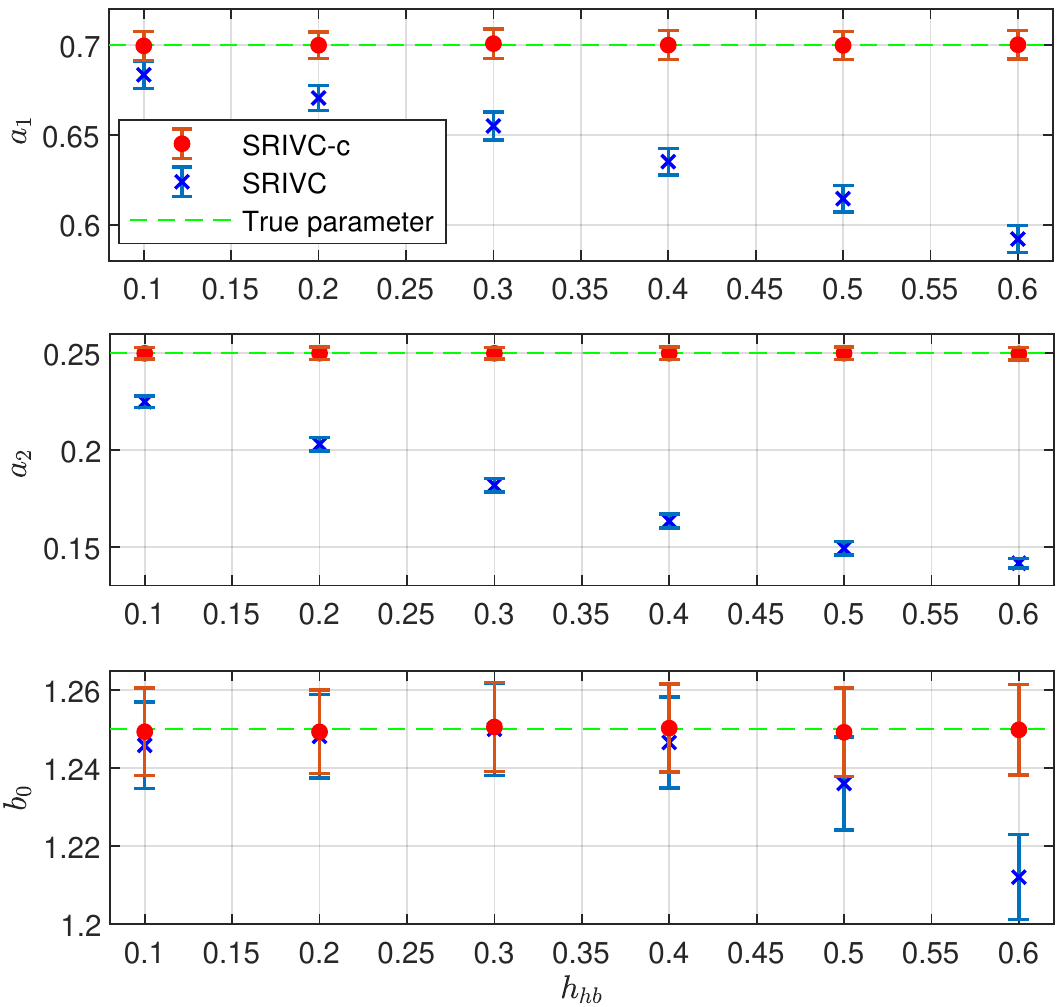}
		\caption{Irregular sampling test. Sample means of each estimated parameter using SRIVC-c (red) and standard SRIVC (blue), with $1$ standard deviation, for different sampling intervals. The true parameter values are in dashed green.}
		\label{fig3}}
\end{figure}

Figure \ref{fig3} shows the mean value of each parameter, with their standard deviation around this value. As expected, the SRIVC-c estimator provides accurate estimates for all sampling period ranges in this study. On the other hand, the SRIVC estimator has a degrading performance as the sampling range increases, which could be attributed to the approximation errors in the prefilter calculations due to incorrect assumptions on the intersampling behaviour.

%%%%%%%%%%%%%%%%%%%%%%%%%%%%%%%%%%%%%%%%%%%%%%%%%%%%%%%%%%%%%%%%%%%%%%%%%%%%%%%%

\section{Conclusions}
\label{sec6}
In this paper, we have derived an algorithm for continuous-time system identification that is consistent for a wide class of input signals that have a known intersample behaviour. This estimator extends the applicability of the standard SRIVC method to continuous-time multisine inputs. This extension also allows estimation using irregularly-sampled data. We put forward a comprehensive analysis of the generic consistency of the proposed estimator for multisine inputs, and extensive simulations have confirmed the theoretical findings and have shown advantages of this estimator over the widely popular SRIVC method. Further research on this topic concerns a variance analysis of this estimator, and theoretical guarantees for irregular sampling schemes.

\begin{ack}            
This work was partially supported by the Swedish Research Council under contract number 2016-06079 (NewLEADS) and by the Australian government Research Training Program (RTP) scholarship.	
\end{ack}

\section{Appendix}
\begin{lem}\hspace{-0.2cm}\textbf{.}
\label{lemma1}
Consider $u(t)$ as in \eqref{inputmultisine}, and $\mathbf{v}_f(t_k)$ and $\mathbf{u}_{du}(t)$ as defined in \eqref{Vtk} and \eqref{Udu} respectively. Under Assumption \ref{assumption2}, the matrix $\bm{\Psi}$ defined in \eqref{Psidef} is equal to zero.
\end{lem}

\begin{pf*}{Proof.}
From the definition of $\mathbf{v}_f(t_k)$, we directly obtain that all the entries $\bm{\Psi}_{il}$ of $\bm{\Psi}$, with \mbox{$l>n+1$}, are equal to zero. For the other entries, we see that an arbitrary entry of this matrix is of the form
\begin{equation}
\label{expec1}
\bm{\Psi}_{il} = \overline{\mathbb{E}}\left\{ \left\{\frac{p^{n+m+1-i}u(t)}{A_j^2(p)}\right\}_{t_k} \frac{p^{l}}{A_j(p)}v(t_k) \right\}.
\end{equation}
If we define $\{g_i(t)\}_{t\geq 0}$ as the inverse Laplace transform of $s^{n+m+1-i}A_j^{-2}(s)$, the first term in the expectation in~\eqref{expec1} can be written as
\begin{equation}
\left\{\frac{p^{n+m+1-i}u(t)}{A_j^2(p)}\right\}_{t_k} = \int_0^{t_k} g_i(t_k-\tau) u(\tau)\textnormal{d}\tau. \notag 
\end{equation}
Note that this is a DT signal, as a function of the time measurements $\{t_k\}$. On the other hand, the second term~\eqref{expec1} can be described by
\begin{equation}
\frac{p^{n+1-l}}{A_j(p)}v(t_k) = \sum_{r=1}^k \beta_{k-r,l} v(t_r). \notag 
\end{equation}
where $\{\beta_{j,l}\}_{j=0}^{k-1}$ are the first $k$ values of the impulse response of the FOH DT equivalent of $p^{l}A_j^{-1}(p)$. So, we compute $\bm{\Psi}_{il}$ as
\begin{align}
\bm{\Psi}_{il} &= \overline{\mathbb{E}}\left\{ \int_0^{t_k} g_i(t_k-\tau) u(\tau)\textnormal{d}\tau \sum_{r=1}^k \beta_{k-r,l} v(t_r) \right\} \notag \\
&=\hspace{-0.05cm} \lim_{N\to \infty}\hspace{-0.05cm} \frac{1}{N} \hspace{-0.08cm}  \sum_{k=1}^N \sum_{r=1}^k \hspace{-0.05cm}  \int_0^{t_k} \hspace{-0.15cm} g_i(t_k\hspace{-0.03cm} -\hspace{-0.03cm} \tau)u(\tau) \beta_{k-r,l} \mathbb{E}\{v(t_r)\} \textnormal{d}\tau \notag \\
&= 0, \notag
\end{align}
where we have used the fact that the disturbance signal has zero mean. \hspace*{\fill} \qed
\end{pf*}

\begin{lem}\hspace{-0.2cm}\textbf{.}
	\label{lemma2}
	Under Assumptions \ref{assumption1} to \ref{assumption5}, with $u(t)$ described as in \eqref{inputmultisine}, the following matrix is generically non-singular with respect to the parameters of the denominator of the model:
	\begin{equation}
	\bar{\bm{\Phi}}:= \overline{\mathbb{E}}\left\{\left[\frac{\mathbf{u}_{du}(t)}{\bar{A}^2(p)}\right]_{t_k} \left[\frac{\mathbf{u}_{du}(t)}{\bar{A}(p)A^*(p)}\right]_{t_k}^\top\right\}. \notag
	\end{equation}
\end{lem}
\begin{pf*}{Proof.}
	Similar to \cite{pan2020consistency}, we follow an analyticity argument. We must first prove that
	\begin{equation}
	\label{phistar}
	\bar{\bm{\Phi}}^*:= \overline{\mathbb{E}}\left\{\left[\frac{\mathbf{u}_{du}(t)}{{A^*}^2(p)}\right]_{t_k} \left[\frac{\mathbf{u}_{du}(t)}{{A^*}^2(p)}\right]_{t_k}^\top\right\}
	\end{equation}
	is positive definite. For this, let $\textbf{z}\in \mathbb{R}^{n+m+1}$. We write
	\begin{equation}
	\textbf{z}^\top \bar{\bm{\Phi}}^* \textbf{z} = \overline{\mathbb{E}}\left\{\left(\left\{\frac{B_{\textbf{z}}(p)}{{A^*}^2(p)}u(t)\right\}_{t_k}\right)^2 \right\} \geq 0. \notag
	\end{equation}
	Since $u(t)$ is a multisine of the form \eqref{inputmultisine}, in steady state we have
	\begin{equation}
	\frac{B_{\textbf{z}}(p)}{{A^*}^2(p)}u(t) = \tilde{\alpha}_0 + \sum_{l=1}^{m_u} \tilde{\alpha}_l \cos(\omega_l t + \tilde{\phi}_l), \notag
	\end{equation}
	where $\tilde{\alpha}_0 = \alpha_0 B_{\textbf{z}}(0)/{A^*}^2(0)$, $\tilde{\alpha}_l = \alpha_l|B_{\textbf{z}}(i\omega_l)/{A^*}^2(i\omega_l)|$, and $\tilde{\phi}_l = \phi_l + \angle B_{\textbf{z}}(i\omega_l)/{A^*}^2(i\omega_l)$. Therefore,
	\begin{subequations}
	\begin{align}
	\textbf{z}^\top \bar{\bm{\Phi}}^* \textbf{z} &= \lim_{N\to \infty} \frac{1}{N} \sum_{k=1}^N \left(\tilde{\alpha}_0 + \sum_{l=1}^{m_u} \tilde{\alpha}_l \cos(\omega_l kh + \tilde{\phi}_l)\right)^2 \notag \\
	\label{cosinesums1}
	&\hspace{-1cm}= \lim_{N\to \infty} \frac{1}{N} \sum_{k=1}^N \Bigg(\tilde{\alpha}_0^2 + 2\tilde{\alpha}_0\sum_{l=1}^{m_u} \tilde{\alpha}_l \cos(\omega_l kh+ \tilde{\phi}_l) \\
	\label{cosinesums2}
	&\hspace{-0.7cm}+\sum_{j,l=1}^{m_u} \tilde{\alpha}_j\tilde{\alpha}_l \cos(\omega_j kh + \tilde{\phi}_j) \cos(\omega_l kh + \tilde{\phi}_l)\Bigg).
	\end{align}
	\end{subequations}
	Recall the formula for a geometric series 
	\begin{align}
	\lim_{N\to \infty} \frac{1}{N}\sum_{k=1}^{N}\cos(\omega k + \phi) &= \textnormal{Re}\left\{\lim_{N\to \infty} \frac{e^{i\phi}}{N} \sum_{k=1}^N e^{i\omega k} \right\} \notag \\
	%&= \textnormal{Re}\left\{\lim_{N\to \infty} \frac{e^{i\phi}}{N} \frac{e^{i\omega}-e^{i\omega(N+1)}}{1-e^{i\omega}}\right\} \notag \\
	&= 0. \notag
	\end{align}
	Using this result, and the identity $\cos(\alpha)\cos(\beta) = [\cos(\alpha+\beta)+\cos(\alpha-\beta)]/2$, the second term in the sum in \eqref{cosinesums1} is zero. Moreover, in \eqref{cosinesums2} the term for $j\neq l$ is a sum of sinusoids whose sum tends to zero as $N$ tends to infinity, while for $j= l$ constants appear. Thus, 
	\begin{equation}
	\label{expectationcomputation}
	\textbf{z}^\top \bar{\bm{\Phi}}^* \textbf{z} = \tilde{\alpha}_0^2 + \frac{1}{2}\sum_{j=1}^{m_u} \tilde{\alpha}_j^2. 
	\end{equation}
	This computation leads to stating that $\textbf{z}^\top \bar{\bm{\Phi}}^* \textbf{z}=0$ occurs if and only if $\tilde{\alpha}_0=\tilde{\alpha}_1 = \cdots = \tilde{\alpha}_{m_u}=0$, which in turn is equivalent to imposing
	\begin{equation}
	\frac{B_{\textbf{z}}(0)}{{A^*}^2(0)} = 0, \hspace{0.12cm} \frac{B_{\textbf{z}}(i\omega_l)}{{A^*}^2(i\omega_l)} = \frac{B_{\textbf{z}}(-i\omega_l)}{{A^*}^2(-i\omega_l)}= 0, \hspace{0.12cm} l = 1,\dots, m_u. \notag
	\end{equation}
	Since $m_u \geq (n+m)/2$, the only rational function that satisfies all of these restrictions is the null transfer function. Thus, $B_{\textbf{z}}(p)=0$ and $\textbf{z}=0$. With this, we have shown that $\bar{\bm{\Phi}}^*$ is positive definite.
	
	We now show that the entries of the matrix $\bar{\bm{\Phi}}$ are real analytic functions of the (real-valued) parameters $(\bar{a}_1,\dots,\bar{a}_n)$ in the domain where $\bar{A}(p)$ is a stable polynomial. We denote this domain as \mbox{$\Omega\subset \mathbb{R}^n$}. The entries of the matrix $\bar{\bm{\Phi}}$ are given by
	\begin{equation}
	\bar{\bm{\Phi}}_{jl}:= \overline{\mathbb{E}}\left\{\left\{\frac{p^{n+m+1-j}u(t)}{\bar{A}^2(p)}\right\}_{t_k}\hspace{-0.05cm} \left\{\frac{p^{n+m+1-l}u(t)}{\bar{A}(p)A^*(p)}\right\}_{t_k}\right\}, \notag 
	\end{equation}
	where $j,l=1,2,\dots,n+m+1$. By computing the expectation similarly to the derivation of \eqref{expectationcomputation}, we find that
	\begin{equation}
	\bar{\bm{\Phi}}_{jl} =\tilde{\alpha}_0^j \tilde{\alpha}_0^l + \frac{1}{2} \sum_{r=1}^{m_u} \tilde{\alpha}_r^j \tilde{\alpha}_r^l \cos(\tilde{\phi}_r^j-\tilde{\phi}_r^l), \notag
	\end{equation}
	where \small
	\begin{align}
	\tilde{\alpha}_0^j &= \begin{cases} 0, & \hspace{-0.2cm} j<n+m+1 \\
\alpha_0, & \hspace{-0.2cm}j= n+m+1\end{cases}, \hspace{0.15cm}
\tilde{\alpha}_0^l = \begin{cases} 0, & \hspace{-0.2cm}l<n+m+1 \\
\alpha_0, & \hspace{-0.2cm}l= n+m+1\end{cases}, \notag \\
\tilde{\alpha}_r^j &= \alpha_r\left|\frac{\omega_r^{n+m+1-j}}{\bar{A}^2(i\omega_r)}\right|, \hspace{0.85cm} \tilde{\phi}_r^j = \phi_r + \angle\left[ \frac{(i\omega_r)^{n+m+1-j}}{\bar{A}^2(i\omega_r)}\right] \notag \\
\tilde{\alpha}_r^l &= \alpha_r \left|\frac{\omega_r^{n+m+1-l}}{\bar{A}(i\omega_r)A^*(i\omega_r)}\right|, \hspace{0.24cm} \tilde{\phi}_r^l = \phi_r + \angle\left[\frac{(i\omega_r)^{n+m+1-l}}{\bar{A}(i\omega_r)A^*(i\omega_r)}\right]. \notag
	\end{align}
    \normalsize
    
    The coefficient $\tilde{\alpha}_r^j$ can be equivalently expressed as
    \begin{align}
    \tilde{\alpha}_r^j &= \frac{\alpha_r \omega_r^{n+m+1-j}}{\textnormal{Re}\{\bar{A}(i\omega_r)\}^2+\textnormal{Im}\{\bar{A}(i\omega_r)\}^2} \notag \\
    \label{ch5:alphar}
    &= \frac{\alpha_r\omega_r^{n+m+1-j}}{\left(\hspace{-0.1cm}1+ \sum\limits_{\substack{1\leq k \leq n \\ k\textnormal{ even}}}
    	 \hspace{-0.2cm}\bar{a}_k \omega_r^k (-1)^{\frac{k}{2}}\hspace{-0.05cm}\right)^{\hspace{-0.1cm}2}\hspace{-0.1cm}+\hspace{-0.05cm}\left(\sum\limits_{\substack{1\leq k \leq n \\ k\textnormal{ odd}}}\hspace{-0.2cm}\bar{a}_k \omega_r^k (-1)^{\frac{k-1}{2}}\hspace{-0.05cm}\right)^{\hspace{-0.1cm}2}}.
    \end{align}
    
    From \eqref{ch5:alphar}, we see that the denominator of $\tilde{\alpha}_r^j$ is a multivariate polynomial in the variables $(\bar{a}_1,\hspace{0.1cm} \dots,\hspace{0.1cm}\bar{a}_n)$, which is strictly positive in $\Omega$, since we know that $\bar{A}(p)$ is a stable polynomial for any $(\bar{a}_1,\hspace{0.1cm} \dots,\hspace{0.1cm}\bar{a}_n)\in \Omega$. This shows that the denominator of $\tilde{\alpha}_r^j$ is real analytic in $\Omega$, and since the quotient of real analytic functions is real analytic as long as the denominator does not vanish \cite[Proposition 2.2.2]{krantz2002primer}, we have that $\tilde{\alpha}_r^j$ is real analytic in $\Omega$.
    
    Similarly, the coefficient $\tilde{\alpha}_r^l$ can be written as
    \begin{equation}
    \tilde{\alpha}_r^l = \frac{\alpha_r\omega_r^{n+m+1-l}}{|A^*(i\omega_r)|\sqrt{\textnormal{Re}\{\bar{A}(i\omega_r)\}^2+\textnormal{Im}\{\bar{A}(i\omega_r)\}^2}}. \notag
  \end{equation}
    Here, the argument of the square root is a real analytic function, and the function $x\rightarrow 1/\sqrt{x}$ is real analytic for $x\in(0,\infty)$. Thus, by \cite[Proposition 2.2.8]{krantz2002primer}, which states that the composition of real analytic functions is real analytic, we conclude that $\tilde{\alpha}_r^l$ is real analytic in $\Omega$.\hspace{-0.05cm}\footnote{Note that the coefficients in $A^*(i\omega_r)$ do not play a role in the analyticity of $\tilde{\alpha}_r^l$, since $\tilde{\alpha}_r^l$ is viewed as a function of $(\bar{a}_1,\hspace{0.1cm} \dots,\hspace{0.1cm}\bar{a}_n)$ only.}
    
    Finally, note that
	\begin{equation}
	\phi_r^j-\phi_r^l = \frac{\pi}{2}(l-j)+\angle A^*(i\omega_r)-\angle \bar{A}(i\omega_r), \notag
	\end{equation}
	which leads to
	\begin{align}
	\cos(\phi_r^j-\phi_r^l) &= \frac{\textnormal{Re}\{\bar{A}(i\omega_r)\}}{|\bar{A}(i\omega_r)|} \cos\left(\frac{\pi}{2}(l-j)+\angle A^*(i\omega_r)\right) \notag \\
	&\hspace{-0.9cm}+ \frac{\textnormal{Im}\{\bar{A}(i\omega_r)\}}{|\bar{A}(i\omega_r)|} \sin\left(\frac{\pi}{2}(l-j)+\angle A^*(i\omega_r)\right). \notag 
	\end{align}
	By the same justification above, $\textnormal{Re}\{\bar{A}(i\omega_r)\}/|\bar{A}(i\omega_r)|$ and $\textnormal{Im}\{\bar{A}(i\omega_r)\}/|\bar{A}(i\omega_r)|$ are real analytic functions for any \mbox{$(\bar{a}_1, \dots,\bar{a}_n)\in \Omega$}. Therefore, $\cos(\phi_r^j-\phi_r^l)$ is real analytic for any $(\bar{a}_1, \dots,\bar{a}_n)\in \Omega$.  Since any function defined by multiplication and addition of real analytic functions is real analytic \cite{krantz2002primer}, we conclude that $\bar{\bm{\Phi}}_{jl}$ is real analytic in the variables $(\bar{a}_1, \dots,\bar{a}_n)$, in the domain $\Omega$.
	
	Thus, for $n=n^*$ and $m\geq m^*$, the matrix $\bar{\bm{\Phi}}$ is generically non-singular with respect to $(\bar{a}_1,\dots,\bar{a}_n)\in \Omega$ by Lemma A2.3 of \cite{soderstrom1983instrumental} and its corollary. If $n\geq n^*$ and $m=m^*$, we note that $\mathbf{a}^*:= (a_1^*,\dots,a_{n^*}^*,0,\dots,0)\in \mathbb{R}^n$ belongs to the boundary of $\Omega$. However, since $\det(\bar{\bm{\Phi}})$ is a real analytic function in $\Omega$, its continuity ensures the existence of a small perturbation vector $\bm{\eta} \in \mathbb{R}^n$ such that $\mathbf{a}^*+\bm{\eta}\in\Omega$ and $\bar{\bm{\Phi}}$ is non-singular when $\mathbf{a}^*+\bm{\eta}$ are the parameters of the model denominator. Hence, generic non-singularity of $\bar{\bm{\Phi}}$ also holds for this case by the same lemma cited above.
	\hspace*{\fill} \qed

\end{pf*}
\balance
\bibliographystyle{plain}        
\bibliography{References}           

\end{document}